\documentclass[sigplan,twocolumn]{acmart}
 
\AtBeginDocument{%
  }

\renewcommand\footnotetextcopyrightpermission[1]{}
\usepackage{enumitem}
\usepackage{multirow}
\usepackage{subcaption}
\usepackage[ruled,vlined,linesnumbered]{algorithm2e}

\newcommand{\name}{FenixWalk}
\AtBeginDocument{%
  }

\begin{document}
\title{{\name}: Adaptive Sampling for Large-Scale Dynamic Graphs on Heterogeneous Architectures}

%%
%% The "author" command and its associated commands are used to define the authors and their affiliations.
% Authors 1--2: Nanjing University
\author{Pinhuan Wang}
\authornote{Both authors contributed equally to this work.}
\affiliation{%
  \institution{Nanjing University}
  \city{Nanjing}
  \country{China}
}
\email{pinhuan.wang02@gmail.com}

\author{Jianian Fu}
\authornotemark[1]
\affiliation{%
  \institution{Nanjing University}
  \city{Nanjing}
  \country{China}
}
\email{f36351010@gmail.com}

\author{Jie Zhang}
\affiliation{%
  \institution{Peking University}
  \city{Beijing}
  \country{China}
}
\email{jiez@pku.edu.cn}

\author{Shixuan Sun}
\affiliation{%
  \institution{Shanghai Jiao Tong University}
  \city{Shanghai}
  \country{China}
}
\email{sunshixuan@sjtu.edu.cn}

\author{Qing Wang}
\affiliation{%
  \institution{Nanjing University}
  \city{Nanjing}
  \country{China}
}
\email{wangqing.cs@nju.edu.cn}

\author{Xiaoliang Wang}
\affiliation{%
  \institution{Nanjing University}
  \city{Nanjing}
  \country{China}
}
\email{waxili@nju.edu.cn}

\author{Chen Tian}
\affiliation{%
  \institution{Nanjing University}
  \city{Nanjing}
  \country{China}
}
\email{tianchen@nju.edu.cn}

\author{Chengying Huan}
\affiliation{%
  \institution{Nanjing University}
  \city{Nanjing}
  \country{China}
}
\email{huanchengying@nju.edu.cn}
\settopmatter{authorsperrow=4}
\renewcommand{\shortauthors}{Wang et al.}

\begin{abstract}
Random walks are fundamental to graph analytics, yet executing them
efficiently on large, evolving graphs remains challenging.
Existing approaches must balance sampling efficiency, update overhead, and memory consumption.
Sampling structures optimized for throughput often incur
substantial memory overhead and costly maintenance, whereas
lightweight alternatives can suffer from high sampling latency
on high-degree vertices or skewed weight distributions.
These challenges become more pronounced on heterogeneous
CPU--GPU platforms, where host memory provides capacity
but GPU memory limits the amount of graph data that can benefit from parallel acceleration.

We present {\name}, a memory-efficient random-walk system
that combines unified in-place sampling with selective
GPU acceleration for large-scale dynamic graphs.
At its core, {\name} employs an in-place Fenwick
representation that supports both rejection sampling
and inverse-transform sampling without duplicating
edge-level weight storage.
This representation enables expected $O(1)$ weight lookup
for rejection sampling, together with $O(\log d)$
prefix sampling and updates.
To exploit heterogeneous hardware efficiently, {\name} maintains the complete graph and its sampling structures
in host memory while selectively placing frequently reused vertex-local samplers on the GPU.
By transferring compact walker states instead of repeatedly migrating graph partitions, the system reduces cross-device communication and makes effective use of limited GPU memory.
Experiments on large real-world graphs and representative
random-walk workloads demonstrate that {\name}
achieves average speedups ranging from
$2.06\times$ to $400.57\times$
over existing random-walk systems.
\end{abstract}

\maketitle

\pagestyle{plain}

\section{Introduction}
\label{sec:intro}

Random walks are a fundamental primitive for graph analytics. 
They are widely used in graph mining and representation learning~\cite{deepwalk,node2vec}, web ranking and personalized search~\cite{pagerank,topic_sensitive_pagerank}, scalable graph neural network training~\cite{graphsaint}, and large-scale recommendation~\cite{pixie,pinsage}. 
For example, DeepWalk~\cite{deepwalk} and node2vec~\cite{node2vec} learn node embeddings from random-walk traces, while PageRank and its variants~\cite{pagerank,topic_sensitive_pagerank} estimate vertex importance from the stationary behavior of random-walk processes.  
Recent graph-learning and recommendation systems further use random walks as a scaling mechanism, supporting compact minibatch construction and efficient item retrieval over massive graphs~\cite{graphsaint,pixie,pinsage}.

Many of these applications run on graphs that evolve over time.
In such graphs, edges are continuously inserted, deleted, or modified, and random walks must be generated on updated snapshots to keep downstream results fresh. 
For example, Wharf~\cite{wharf} shows that reusing stale personalized PageRank walks after graph updates can, in practice, result in a symmetric mean absolute percentage error exceeding $80\%$, indicating that results derived from stale walks can differ substantially from those computed on the updated graph.

Keeping random-walk results fresh on dynamic graphs ultimately relies on transition sampling over updated graph snapshots: given a vertex, the system samples the next vertex from its candidate neighbors according to a transition distribution.
Even on static graphs, this operation can dominate end-to-end execution time; prior work reports that sampling accounts for up to $96.2\%$ of graph-learning training time~\cite{gsampler}. 
Dynamic graphs further complicate this because every edge insertion or deletion changes the transition distribution of the affected vertices.
Therefore, supporting dynamic random walks requires more than updating graph storage: the sampling process must also adapt to updated transition distributions. Maintaining efficient transition sampling under frequent graph updates thus remains a critical \textit{algorithmic challenge}: the sampler must provide high sampling throughput while keeping update and memory costs under control.

\noindent\textbf{Algorithmic Challenge.}
Existing sampling methods expose a three-way trade-off among
\textit{sampling performance, update cost, and memory overhead}.
Alias-based systems~\cite{knightking,skywalker} provide $O(1)$ sampling after preprocessing, but they store auxiliary alias and probability entries and may require rebuilding vertex-local tables after edge or weight updates, incurring substantial update overhead.
Rejection- and reservoir-based systems avoid auxiliary sampling structures and are therefore easy to update~\cite{nextdoor,flowwalker,flexiwalker}, but their per-sample cost can be high: rejection sampling may require many trials under skewed weights, while reservoir sampling scans the adjacency list and is inefficient for high-degree vertices.
Inverse-transform sampling (ITS) supports $O(\log d_v)$ sampling through binary search over a maintained prefix-sum array~\cite{c-saw}, but the array incurs additional memory overhead and can be expensive to maintain under updates.
Bingo~\cite{bingo} improves sampling and update performance for dynamic graphs, but its radix-based decomposition introduces substantial memory overhead.
Consequently, existing designs still face a tension among efficient sampling, inexpensive updates, and low memory overhead.

This trade-off motivates a closer look at rejection sampling and ITS.
Rejection sampling is attractive for dynamic graphs because each proposal only requires the weight of a uniformly sampled edge position, but its efficiency degrades as the acceptance probability decreases.
ITS, in contrast, is robust to weight skew, but its conventional prefix-sum array is costly to maintain under updates.
A Fenwick tree provides a natural alternative, supporting both ITS sampling and weight updates in $O(\log d)$ time.
Our key observation is that \textit{the Fenwick representation can also provide efficient point-weight lookup for rejection sampling}.
Although recovering an arbitrary edge weight from a Fenwick tree can take $O(\log d)$ time in the worst case, the cost depends on the number of trailing zero bits of the queried index.
Since rejection sampling proposes edge positions uniformly, this lookup requires fewer than two Fenwick-entry accesses on average.
Thus, a single Fenwick representation can support both sampling strategies without retaining a separate weight array.

Based on this observation, our core idea is to \textit{directly encode each vertex's edge weights in an in-place Fenwick representation}.
For a vertex of degree $d$, {\name} transforms the original $d$ weight-sized slots into a Fenwick tree of the same length.
The representation supports $O(\log d)$ prefix sampling and updates for ITS, while providing expected $O(1)$ weight lookup for rejection sampling.
{\name} can therefore choose between rejection sampling and ITS at runtime according to the degree and weight distribution of each vertex, without changing or duplicating the underlying sampling state.
This in-place representation combines the robustness of ITS with the expected efficiency of rejection sampling while using only the original edge-level weight storage.

\noindent\textbf{System Challenge.} The proposed in-place sampler alone addresses the algorithmic challenge, but deploying it at large scale introduces a system challenge.
Modern graphs can contain billions of edges, and a random-walk engine must store the graph topology, edge weights, sampling state, and active walker state.
Host DRAM can hold such graphs, but CPU-only execution provides limited throughput for the massive number of independent walk steps.
GPUs offer high bandwidth and parallelism, but device memory is much smaller than host memory.
Even with a compact sampler, the entire dynamic graph may not fit on the GPU.
A natural response is hybrid CPU--GPU execution, but existing designs are not well matched to random walks.
Systems that repeatedly transfer graph partitions across the CPU--GPU boundary can move far more data than each walk step actually consumes~\cite{lighttraffic}.
A random-walk step touches only the current vertex's local distribution, and the next vertex is known only after sampling.
The system challenge is therefore to exploit GPU parallelism while treating GPU memory as a limited execution resource.

To address above challenges, we present \textbf{\name}, a hybrid CPU--GPU random-walk engine for large dynamic graphs. 
{\name} combines the in-place dynamic sampler with selective GPU residency.
It keeps the complete dynamic graph and all samplers in CPU memory while placing only high-reuse vertex-local samplers on the GPU.
Walkers located at GPU-resident vertices perform transition sampling on the GPU; walkers at nonresident vertices are processed by CPU worker threads.
During normal execution, {\name} transfers active walker states rather than repeatedly moving adjacency lists or sampling structures.
The resident set is refreshed at epoch granularity using visit-frequency estimates and memory cost, so GPU memory is spent on vertices whose expected reuse justifies their device footprint.

The in-place sampler and the hybrid CPU--GPU engine reinforce each other.
Because each sampler uses only the original weight slots, more useful vertices can remain resident under the same GPU memory budget.
This combination lets {\name} support dynamic graphs while using CPU memory for capacity and the GPU for high-reuse sampling.

This paper makes the following contributions:

\begin{itemize}[leftmargin=*]
\item We propose a dynamic transition-sampling method that unifies rejection sampling and inverse-transform sampling in a single sampler and selects the more efficient strategy for each vertex at runtime (Section~\ref{sec:algo}).

\item We show that a Fenwick representation can serve both sampling strategies. It supports $O(\log d)$ prefix sampling and updates for ITS, while providing expected $O(1)$ point-weight lookup for rejection sampling, using only the original $d$ weight-sized slots (Section~\ref{sec:algo}).

\item We build a hybrid CPU--GPU random-walk engine that keeps the dynamic graph in host DRAM and places only high-reuse vertex-local samplers on the GPU, avoiding repeated migration of large adjacency lists or sampling structures (Section~\ref{sec:arch}).

\item We evaluate {\name} on large real-world graphs and representative random-walk workloads under dynamic updates. By combining compact in-place samplers with selective GPU residency, {\name} achieves $2.06\times$ to $400.57\times$ average walk-execution speedups over existing random-walk systems (Section~\ref{sec:evaluation}).
\end{itemize}

\section{Background and Motivation}
\label{sec:background}

\subsection{Random Walks and Transition Sampling}
\label{sec:bg-random-walk}

Let \(G=(V,E)\) be a graph, which can be directed or undirected. For a vertex \(v\in V\), let \(N(v)\) denote the candidate neighbors that can be visited from \(v\), and let \(d_v=|N(v)|\). In a directed graph, \(N(v)\)
contains the outgoing neighbors of \(v\). In an undirected graph, each undirected edge can be viewed as two directed arcs, so the same notation applies.

A random walk is a sequence of vertices $X_0, X_1, \ldots, X_L$, where each next vertex is sampled from the candidates of the current walk state. Random walks are widely used in graph analytics. For example, DeepWalk~\cite{deepwalk} uses random walks to generate vertex sequences for representation learning, while node2vec~\cite{node2vec} extends this idea with biased walks that control how the walk explores local and global neighborhoods.

The next-hop sampling in random walks can be uniform or weighted. We use the weighted form as the common notation: each candidate edge \((v,u)\) is assigned a nonnegative transition weight \(w_{v,u}\), and the next vertex is sampled according to
\begin{equation*}
  \Pr[X_{i+1}=u \mid X_i=v]
  =
  \frac{w_{v,u}}{\sum_{x\in N(v)} w_{v,x}}.
\end{equation*}
The uniform random walk is recovered by setting \(w_{v,u}=1\) for all \(u\in N(v)\). Many random-walk methods~\cite{pagerank,deepwalk} use weights that depend only on the current vertex and candidate neighbor, while methods~\cite{node2vec,metapath} such as node2vec can use weights that also depend on the previous walk state.

\subsection{Random Walks on Dynamic Graphs}
\label{sec:bg-dynamic-graphs}

A dynamic graph evolves through a sequence of edge and vertex updates, including insertions, deletions, and weight or attribute changes. For random-walk execution, vertex updates can be handled through edge changes.
We model the dynamic graph at time \(t\) as a
snapshot \(G_t=(V_t,E_t)\), and random walks may be executed on selected snapshots to keep downstream applications up to date.

Preparing the next selected snapshot for random-walk execution requires not only updating the graph representation, which is typically handled by the dynamic-graph container, but also updating any auxiliary sampling structures used by the sampler to ensure correct transition sampling. This paper, {\name}, focuses on the latter requirement: auxiliary sampling structures
must be maintained efficiently while still supporting fast sampling, so that random-walk execution can resume quickly after graph updates.

This setting differs from \emph{dynamic random walks} studied by recent systems~\cite{flowwalker,flexiwalker}. In dynamic random walks, the graph topology may be fixed, while the transition distribution is computed at runtime from user-defined walk logic or state, as discussed in Section~\ref{sec:bg-random-walk}.

\subsection{Sampling Methods}
\label{sec:bg-sampling}

From a systems perspective, the random-walk variants discussed above share the same core operation: given the current walk state, sample one candidate neighbor according to the corresponding transition distribution. Thus, each walk step reduces to transition sampling over a candidate set, and a random-walk engine must perform this operation repeatedly for many concurrent walkers.

\begin{figure}
\centerline{\includegraphics[width=.95\linewidth]{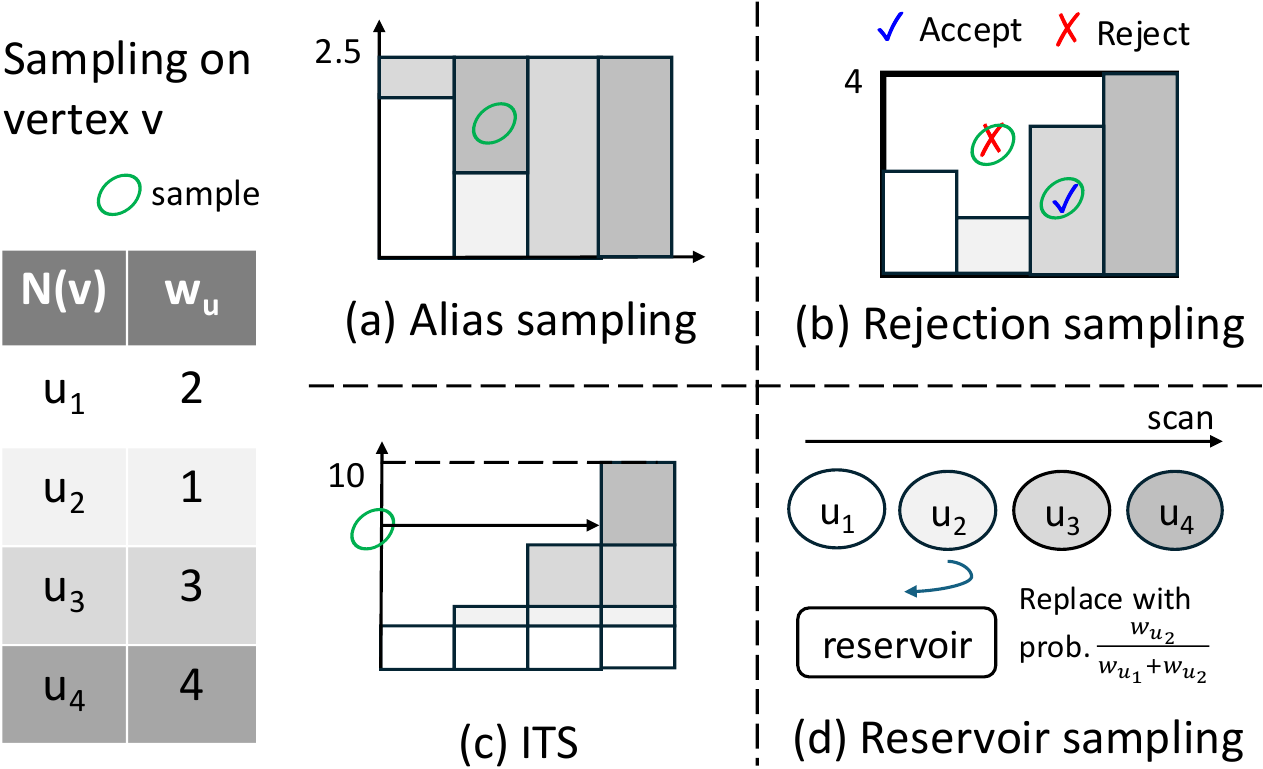}}
    \caption{Examples of four representative sampling methods.}
    \vspace{-2em}
    \label{fig:sampling-methods}
\end{figure}

For dynamic graphs, the sampling method must balance three requirements: high sampling throughput, efficient updates after graph changes, and low memory overhead. These requirements are often in tension. A sampler with fast sampling may require expensive preprocessing or large auxiliary structures, while an update-friendly sampler may have slower per-step sampling. We illustrate four representative sampling methods in Figure~\ref{fig:sampling-methods} and summarize them below.

\noindent\textbf{Alias Sampling.}
As shown in Figure~\ref{fig:sampling-methods}(a), alias methods partition the weighted outgoing edges of vertex \(v\) into \(d_v\) buckets with the same total weight, where each bucket stores at most two candidate edges. Sampling first selects a bucket uniformly at random and then chooses between the stored candidates. After \(O(d_v)\) preprocessing, alias tables support \(O(1)\) weighted sampling, making them attractive for static graphs.
However, alias tables are costly to maintain under updates: Any change to the outgoing edges may require rebuilding the affected vertex's alias structure. They also store extra probability and alias entries, increasing the overall memory footprint. These costs make alias sampling less suitable for large, dynamically changing graphs.

\noindent\textbf{Rejection Sampling.}
Rejection sampling repeatedly selects a candidate neighbor uniformly at random and accepts it with probability proportional to its weight. Figure~\ref{fig:sampling-methods}(b) shows examples of accepted and rejected candidates. Its main advantage is that it avoids auxiliary sampling structures and only relies on the current edge weights in a dynamic graph. However, its runtime depends on the acceptance rate and can degrade to \(O(d_v)\) under skewed weight distributions.

\noindent\textbf{Inverse Transform Sampling.}
Inverse transform sampling (ITS) draws a random value
$r \in \left[0,\sum_{u\in N(v)} w_{v,u}\right)$
and returns the first edge whose prefix sum exceeds \(r\). If a prefix-sum array is maintained, as shown in Figure~\ref{fig:sampling-methods}(c), sampling can be reduced to \(O(\log d_v)\) via binary search. However, this prefix-sum array introduces extra memory overhead and must be updated as the graph changes.

\noindent\textbf{Reservoir Sampling.}
Reservoir sampling processes the neighbors of \(v\) in a streaming manner, as shown in Figure~\ref{fig:sampling-methods}(d). It scans the adjacency list once and maintains a single current sample, or reservoir. When an edge \((v,u)\) with weight \(w_{v,u}\) is visited, the reservoir is updated according to the cumulative weight seen so far, so that the final selected edge is sampled with probability proportional to its weight. Reservoir sampling avoids precomputed sampling structures and can be effective when transition weights are computed on the fly. However, each sample requires scanning all \(d_v\) neighbors, resulting in \(O(d_v)\) time per step. This makes it inefficient for high-degree vertices and large graphs.

\noindent\textbf{Other Methods.}
Beyond the methods above, Bingo~\cite{bingo} uses binary decomposition of edge weights to support \(O(1)\) sampling and \(O(\log W)\) updates on dynamic graphs, where \(W\) denotes the maximum edge weight. However, this design requires \(O(d_v \log W)\) additional space for a vertex \(v\), which substantially increases the memory footprints.

%\vspace{-1em}
\subsection{Limitations of Prior Work and Motivation}
\label{sec:motivation}

As discussed in Section~\ref{sec:bg-sampling}, existing random-walk samplers do not simultaneously balance sampling speed, update cost, and memory footprint. Alias sampling provides $O(1)$ sampling, but its tables are costly to update and require extra storage. Bingo also supports $O(1)$ sampling and efficient updates, but requires $O(d_v \log W)$ extra space for vertex $v$. ITS provides an $O(\log d_v)$ sampling bound, but requires a maintained prefix structure. Rejection and reservoir sampling avoid auxiliary structures and are easy to update, but their cost can degrade to $O(d_v)$ on high-degree vertices.

This trade-off becomes more critical on GPUs. Random walks expose massive parallelism, but large dynamic graphs often exceed GPU memory once graph, sampling structures, and active walk states are included. Even memory-efficient systems~\cite{flowwalker,flexiwalker} cannot fully avoid this. Hybrid CPU--GPU execution is therefore attractive: the CPU provides large DRAM capacity, while the GPU accelerates parallel walks. However, out-of-GPU systems such as LightTraffic move graph partitions and walk states between CPU and GPU memory. This transfer-driven execution can be inefficient for random walks, whose low arithmetic intensity makes CPU--GPU traffic and graph loading potential bottlenecks~\cite{lighttraffic}.

These algorithmic and systemic observations motivate {\name}. We need a sampler that supports efficient updates with minimal persistent edge-level memory, allowing more vertices to fit in GPU memory. We also need a hybrid execution model that exploits GPU parallelism and host-memory capacity without repeatedly moving large adjacency lists or sampling structures. {\name} addresses these requirements with an in-place dynamic sampler (Section~\ref{sec:algo}) and a CPU--GPU execution architecture (Section~\ref{sec:arch}).

\section{In-Place Sampling for Dynamic Graphs}
\label{sec:algo}

\noindent\textbf{Observation.}
The sampling methods in Section~\ref{sec:bg-sampling} expose different trade-offs.
Rejection and reservoir sampling require no auxiliary sampling structures and are therefore easy to maintain under updates.
Between them, rejection sampling is more attractive for random walks:
when transition weights are not highly skewed, it requires only a constant expected number of proposals, whereas reservoir sampling
always scans the full adjacency list in $O(d_v)$ time per sample.

Inverse transform sampling (ITS), in contrast, provides a robust $O(\log d_v)$ sampling by searching cumulative weights.
Rejection sampling and ITS therefore offer complementary strengths across different weight distributions.
However, ITS requires prefix-sum information, while rejection sampling requires direct access to individual edge weights. A straightforward implementation supporting both access patterns may therefore retain the raw edge weights together with a separate prefix-sum structure, increasing storage.

\noindent\textbf{Our Goal.}
We seek a single vertex-local representation that supports both sampling
strategies without duplicating edge-level weight state. For a vertex $v$
with degree $d_v$, the representation should

\begin{itemize}[leftmargin=*]
    \item use only $d_v$ weight-sized slots;
    \item support expected $O(1)$ weight lookup for uniformly sampled positions, preserving efficient rejection sampling; and
    \item support $O(\log d_v)$ prefix selection and updates, preserving the robustness of ITS.
\end{itemize}

We next show that the Fenwick-tree~\cite{fenwick} representation satisfies these requirements.

\subsection{In-Place Fenwick Encoding}
\label{sec:inplace-overlay}

\noindent\textbf{Key Idea.}
For a vertex $v$ with degree $d=d_v$, let
$W=(w_1,w_2,\ldots,w_d)$ denote its transition weights.
Instead of storing $W$ together with a separate sampling structure,
{\name} transforms these $d$ weight-sized slots directly into a Fenwick-tree
representation $F$ of the same length.

For position $i$, let $\mathrm{lowbit}(i)$ denote the least significant
set bit of $i$. The value stored at position $i$ is

\begin{equation}
\label{eq:fenwick-layout}
F[i]
=
\sum_{j=i-\mathrm{lowbit}(i)+1}^{i} w_j .
\end{equation}

Thus, every array position stores the sum of the standard Fenwick range
ending at that position. No additional edge-level array is allocated.
The transformation from the original weights to this representation can
be performed in $O(d)$ time using the linear Fenwick-tree construction.

An important property of this layout is that the amount of information aggregated at position $i$ depends on $\mathrm{lowbit}(i)$. For example,
every odd position satisfies $F[i]=w_i$, while positions divisible by
larger powers of two summarize correspondingly larger ranges. As we show below, this structure makes point-weight recovery particularly
efficient when queried positions are uniformly distributed.

\begin{figure}[t]
    \centering
    \includegraphics[width=.95\linewidth]{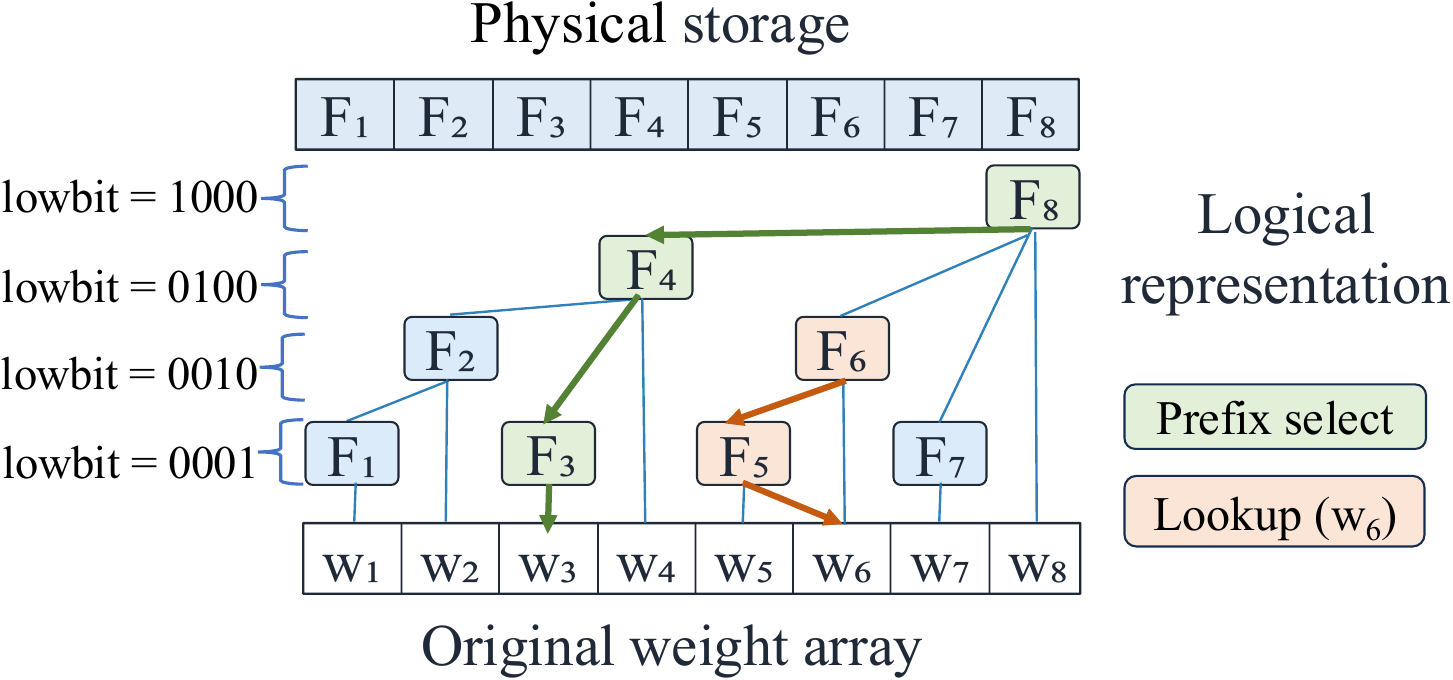}
    \vspace{-.5em}
    \caption{Fenwick encoding in {\name}. Each $F[i]$ covers the range ending at $i$ with length $\mathrm{lowbit}(i)$. Highlighted paths illustrate prefix selection and point-weight lookup.}
    \label{fig:inplace-overlay}
    \vspace{-1.5em}
\end{figure}

Figure~\ref{fig:inplace-overlay} illustrates the representation.
The array supports two operations required during walk execution:
prefix selection for ITS and point-weight lookup for rejection sampling.

\subsection{Sampling and Update Operations}
\label{sec:operations}

\noindent\textbf{Sampling Primitives.}
Algorithm~\ref{alg:adaptive-sampling} shows how the in-place representation supports weight lookup and prefix selection, enabling {\name}'s adaptive sampling.

For ITS, \textsc{PrefixSelect} draws a random value
$r\in[0,W_v^{\Sigma})$ and performs a Fenwick-tree binary search
to find the first position whose cumulative weight exceeds $r$.

For rejection sampling, a candidate position $i$ is chosen uniformly
from $\{1,\ldots,d_v\}$. \textsc{WeightLookup} recovers the corresponding
original weight from the Fenwick representation. Starting from $F[i]$,
it subtracts entries that cover the preceding positions in the range represented by $F[i]$.

\noindent\textbf{Adaptive Sampling.}
The representation allows {\name} to choose between rejection
sampling and ITS without changing the underlying sampling structure.
For each sampling request at vertex $v$, {\name} estimates their costs as

\begin{equation}
\label{eq:choose}
T_{\mathrm{rej}}(v)
=
c_{\mathrm{lookup}}
\cdot
\frac{d_v W_v^{\max}}{W_v^{\Sigma}},
\qquad
T_{\mathrm{its}}(v)
=c_{\mathrm{find}}\log d_v,
\end{equation}

where $W_v^{\Sigma}$ is the total transition weight,
$W_v^{\max}$ is an upper bound on edge weight, and
$c_{\mathrm{lookup}}$ and $c_{\mathrm{find}}$ capture the implementation
costs of the two primitives. The rejection estimate follows from an
acceptance probability of $W_v^{\Sigma}/(d_vW_v^{\max})$.

During walk execution, {\name} compares the two estimates and uses the
cheaper strategy for the current vertex. When weights are relatively
uniform, rejection sampling avoids the logarithmic prefix search.
When the distribution is highly skewed, {\name} switches to ITS and
avoids repeated rejected proposals.

\SetKwProg{Fn}{Function}{:}{}

\begin{algorithm}[t]
\caption{{\name's} Adaptive sampling}
\label{alg:adaptive-sampling}
%\footnotesize
%\SetAlgoNlRelativeSize{-1}
\SetInd{0.4em}{0.8em}

\KwIn{Vertex $v$; random number generator $\mathcal{R}$;
      edge position $i$; cost ratio
      $\rho=c_{\mathrm{lookup}}/c_{\mathrm{find}}$}
\KwOut{Sampled edge position or true edge weight}
\vspace{-.15em}

\Fn{\textsc{WeightLookup}$(v,i)$}{
  $x,z,j \leftarrow F_v[i],i-\mathrm{lowbit}(i),i-1$\;
  \lWhile{$j\neq z$}{
    $x,j\leftarrow x-F_v[j],j-\mathrm{lowbit}(j)$
  }
  \KwRet $x$\;
}
\vspace{-.25em}

\Fn{\textsc{PrefixSelect}$(v,\mathcal{R})$}{
  $r\leftarrow\mathrm{Uniform}(\mathcal{R},0,W_v^\Sigma)$\;
  $p,s,b\leftarrow0,0,2^{\lfloor\log_2 d_v\rfloor}$\;
  \While{$b>0$}{
    $q,b\leftarrow p+b,\lfloor b/2\rfloor$\;
    \lIf{$q\le d_v$ \textbf{and} $s+F_v[q]\le r$}{
      $p,s\leftarrow q,s+F_v[q]$
    }
  }
  \KwRet $p+1$\;
}
\vspace{-.15em}

\Fn{\textsc{RejectionSample}$(v,\mathcal{R})$}{
  \Repeat{$r<w$}{
    $i\leftarrow\mathrm{UniformInt}(\mathcal{R},1,d_v)$\;
    $w,r\leftarrow \textsc{WeightLookup}(v,i),
      \mathrm{Uniform}(\mathcal{R},0,W_v^{\max})$\;
  }
  \KwRet $i$\;
}
\vspace{-.25em}

\Fn{\textsc{AdaptiveSample}$(v,\mathcal{R})$}{
  \If{$\rho\frac{d_vW_v^{\max}}{W_v^\Sigma}\le\log d_v$}{
    \KwRet $\textsc{RejectionSample}(v,\mathcal{R})$\;
  }
  \KwRet $\textsc{PrefixSelect}(v,\mathcal{R})$\;
}
\vspace{-.25em}
\end{algorithm}

\noindent\textbf{Weight Updates.}
For an additive update $\Delta$ to position $i$, {\name} applies the Fenwick-tree point update:

$
F[j] \leftarrow F[j]+\Delta,
\;\;\;\;
j=i,\;
i+\mathrm{lowbit}(i),\ldots,d.
$

For a set-value update from $w_i$ to $w_i'$, {\name} first obtains
$w_i$ using \textsc{WeightLookup}, computes
$\Delta=w_i'-w_i$, and then performs the same Fenwick point update.

{\name} also maintains $W_v^{\Sigma}$ using the same update delta.
This requires only constant-size metadata per vertex.

\noindent\textbf{Insertions and Deletions.}
The dynamic graph container appends newly inserted edges to the vertex-local adjacency array. Suppose a new edge with weight $w_i$ is
appended at position $i=d+1$. Its Fenwick entry can be initialized as

\begin{equation}
\label{eq:fenwick-append}
F[i]
=
w_i
+
\mathrm{Prefix}(i-1)
-
\mathrm{Prefix}(i-\mathrm{lowbit}(i)),
\end{equation}

where $\mathrm{Prefix}(x)$ denotes the prefix sum through $x$.
%Thus, appending a new position requires $O(\log d)$ time without rebuilding the existing representation.

A deletion is implemented lazily by setting the corresponding edge
weight to zero through a set-value update. The zero-weight entry remains in the array until the dynamic graph container performs its periodic compaction.

When the adjacency array is rebuilt, {\name} first recovers the edge weights, removes deleted entries, and constructs a new Fenwick representation for the compacted array. Further details of dynamic graph maintenance
are presented in Section~\ref{subsec:dynamic-graph-support}.

%\noindent\textbf{Maximum Weight Maintenance.}
%{\name} maintains $W_v^{\max}$ as an upper bound on the maximum transition weight of vertex $v$. When an insertion or weight increase produces a larger value, the bound is updated immediately. If an edge attaining the current maximum is decreased or deleted, {\name} retains the previous value instead of maintaining an additional dynamic maximum structure.

%A stale upper bound does not affect the correctness of rejection sampling, because it only enlarges the rejection envelope, although it may temporarily reduce the acceptance probability. During adjacency-array rebuilding, {\name} recomputes the exact maximum weight and restores the tight bound.

\subsection{Complexity Analysis}
\label{subsec:complexity}

We analyze the vertex-local sampling structure independently.
Consider a vertex $v$ whose current Fenwick array contains $d$
edge positions, including zero-weight entries retained before compaction, and assume $W_v^{\Sigma}>0$.

\noindent\textbf{Sampling.}
Prefix selection examines at most one entry per tree level, so \textsc{PrefixSelect} takes $O(\log d)$ time.

For \textsc{WeightLookup} at position $i$, let $\nu_2(i)$ denote the
number of trailing zero bits of $i$.
Algorithm~\ref{alg:adaptive-sampling} performs one initial access to $F_v[i]$ followed by exactly $\nu_2(i)$ subtraction steps.
Rejection sampling proposes a position
$I$ uniformly from $\{1,\ldots,d\}$.
Therefore,
\begin{equation}
\label{eq:less_than_2}
\mathbb{E}[1+\nu_2(I)]
=
1+
\sum_{k\ge1}
\frac{\lfloor d/2^k\rfloor}{d}
<2.
\end{equation}
Hence, each rejection proposal requires expected $O(1)$ Fenwick accesses.
Since its acceptance probability is
$\frac{W_v^{\Sigma}}{dW_v^{\max}}$,
\textsc{RejectionSample} takes expected $O(\frac{dW_v^{\max}}{W_v^{\Sigma}})$ time.

{\name} selects between rejection sampling and prefix selection
using Equation~\ref{eq:choose}.
Ignoring the constant calibration factors captured by
$c_{\mathrm{lookup}}$ and $c_{\mathrm{find}}$, the expected cost of
adaptive sampling is therefore
$
O~(
\min\{
\frac{dW_v^{\max}}{W_v^{\Sigma}},
\log d
\}
).
$
Thus, {\name} preserves expected constant-time sampling when rejection is effective, while retaining an expected $O(\log d)$ bound under skewed weight distributions.

\noindent\textbf{Updates and Rebuilding.}
An additive update modifies at most $O(\log d)$ Fenwick entries.
A set-value update first recovers the old weight and then applies the corresponding additive update, giving $O(\log d)$ time.
Appending an edge using Equation~\ref{eq:fenwick-append} requires $O(\log d)$ time, and lazy deletion is a set-value update, which also takes $O(\log d)$ time.

When compaction rebuilds a vertex-local array, recovering all original weights takes \(O(d)\) work, since the average lookup cost over all positions is less than two as shown in Equation~\ref{eq:less_than_2}. Removing zero-weight entries and constructing the new Fenwick representation are also linear, so the complete rebuild takes \(O(d)\) work.

\noindent\textbf{Construction and Space.}
The initial representation of a vertex can be constructed in $O(d)$ time.
Over a graph with $m$ edges, initialization therefore
requires $O(m)$ work.

For a vertex with $d$ edge positions, the sampling representation
uses one weight-sized value per position and $O(1)$ additional metadata, including $W_v^{\Sigma}$ and $W_v^{\max}$.
Thus, over the graph, the sampler requires $m$ weight-sized edge-level entries plus $O(|V|)$ metadata.
In contrast, retaining both the raw weights and a separate data structure requires $2m$ edge-level weight entries.
The spare capacity maintained by the dynamic graph container is orthogonal to this saving: {\name} does not introduce a second sampling array.

\noindent\textbf{Discussion.}
The key property exploited by {\name} is the match between the Fenwick representation and the access pattern of rejection sampling. For uniformly sampled positions, point queries require fewer than two entry accesses on average. This allows a single encoding to support both efficient rejection sampling and ITS, while retaining logarithmic updates and avoiding a second edge-level array.

\begin{figure}
    \centering
        \includegraphics[width=\linewidth]{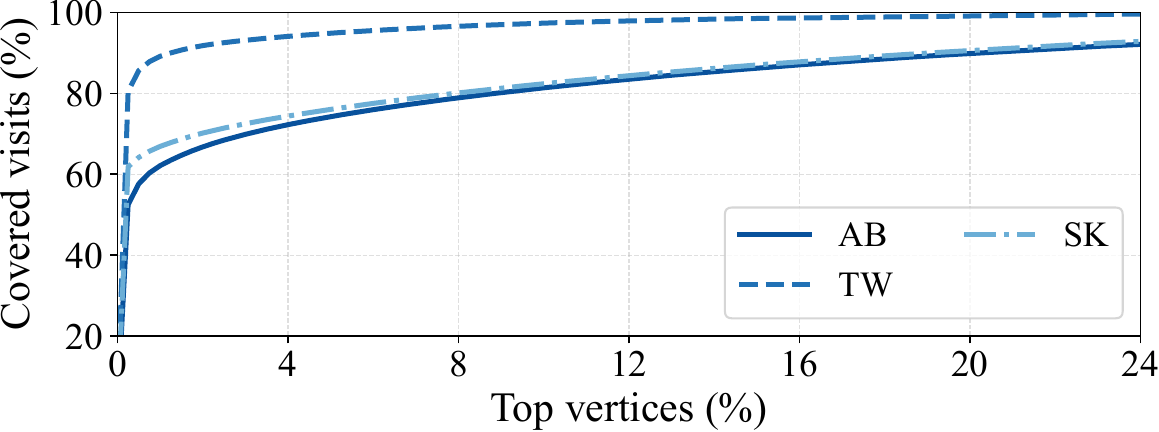}
    \caption{Cumulative coverage of sampling visits by the top-visited vertices on datasets AB, TW, and SK.}
    \label{fig:skew_visits}
    \vspace{-1.5em}
\end{figure}

\section{Hybrid CPU--GPU Architecture}
\label{sec:arch}

Section~\ref{sec:algo} introduces an in-place sampler that avoids a separate sampling structure. This reduction is necessary but not sufficient for large graphs. {\name} therefore treats GPU memory as a limited execution resource rather than as a place to store the whole graph.

\subsection{Architecture Overview}
\label{subsec:arch_overview}

%Section~\ref{sec:algo} introduces an in-place sampler that avoids separate sampling structure. This reduction is necessary but not sufficient for large graphs. {\name} therefore treats GPU memory as a limited execution resource rather than as a place to store the whole graph.

\noindent\textbf{Observation.} The key observation is that random-walk visits are not uniform. 
Figure~\ref{fig:skew_visits} shows the cumulative fraction of sampling visits covered by the top-visited vertices during a DeepWalk execution. 
Across all datasets, a small fraction of vertices accounts for most steps: the top $10\%$ cover more than $80\%$ of visits. The skew is particularly strong on TW, where the top $1\%$ of vertices covers about $90\%$ of visits. This concentration largely reflects structural skew in the graph.

This observation suggests that GPU memory should be used for reusable vertex-local sampling structures rather than for uniformly storing graph data. Instead of treating GPU memory as a cache for graph partitions, {\name} treats it as a limited resource for high-reuse sampling states. The host keeps the complete dynamic graph and manages walk execution, while the GPU persistently stores only the vertex-local sampling states of selected vertices.

%Over an execution epoch, a small subset of vertices can account for a large fraction of sampling steps, due to structural skew, biased transition rules, and workload-specific access patterns. {\name} uses this reuse signal to decide which vertex-local states should reside on the GPU. The CPU keeps the complete dynamic graph and manages walk execution, while the GPU stores only the vertex-local sampling arrays for selected vertices.

This design differs from transfer-driven CPU--GPU execution. 
Systems such as LightTraffic~\cite{lighttraffic} reduce CPU--GPU traffic by coordinating graph partitions and walk states, but their execution still centers on moving graph data needed by active walkers. 
For fine-grained random-walk steps, this can be inefficient because each step uses only a small part of the loaded data, and future accesses are difficult to predict. 
LightTraffic's evaluation likewise shows substantial CPU--GPU I/O overhead on large graphs~\cite{lighttraffic}. 
In addition, such designs are less suitable for biased dynamic sampling, where the system must maintain and access updatable sampling states rather than only traverse adjacency data. {\name} instead keeps selected reusable sampling states persistently on the GPU and transfers only walker states during execution.

\begin{figure}[t]
    \centering
    \includegraphics[width=.85\linewidth]{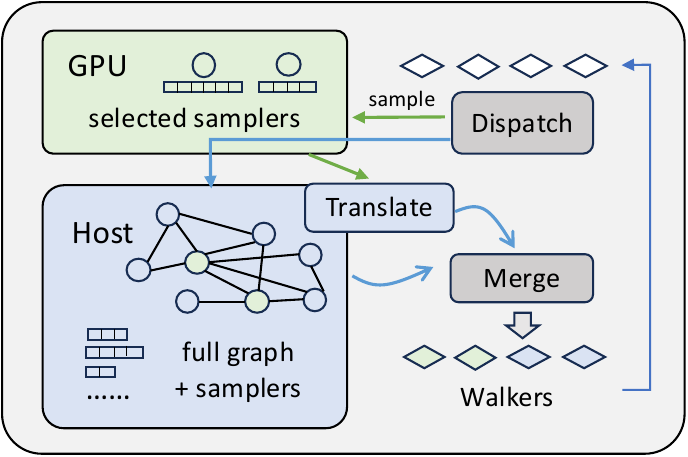}
    \caption{Overview of \name's hybrid CPU--GPU architecture. The left side shows the memory layout, and the right side shows the execution flow. }
    \label{fig:arch}
    \vspace{-1em}
\end{figure}

Figure~\ref{fig:arch} gives an overview of \name's hybrid CPU--GPU architecture. The host keeps the graph together with all samplers introduced in Section~\ref{sec:algo}, while the GPU stores only a selected subset of samplers. During walk execution, walkers are dispatched according to the residency of their current vertices. Steps on GPU-resident vertices are processed by GPU sampling kernels, which return local neighbor positions that are translated on the host into global vertex identifiers. 
Steps on remaining vertices are served directly from the host-resident graph, and the two streams are merged into the final walk results. This organization allows CPU and GPU execution to proceed concurrently while avoiding repeated movement of full adjacency lists or graph partitions.

\subsection{Device-Resident Vertex Selection}
\label{subsec:resident_selection}

{\name} selects a set of vertex sampler arrays to keep on the GPU during preprocessing. This selection is important because an effective residency strategy can improve GPU utilization. The selection is guided by two factors: how often each vertex is expected to be visited, and how much memory is required to keep its sampler array resident. This formulation avoids selecting vertices by visit count alone,
which can waste device memory on very high-degree vertices whose sampler arrays are expensive to keep on the GPU.

The initial visit-frequency estimate for each vertex can be obtained from a small CPU warm-up run.
Let \(\widehat{f}(v)\) denote the estimated number of visits to vertex \(v\) in the upcoming epoch. 
Let \(m_{\mathrm{gpu}}(v)\) denote the GPU memory
footprint of keeping \(v\)'s in-place sampling structure on the device. 
Let \(\Delta t_{\mathrm{gpu}}(v)\) denote the estimated per-visit latency reduction from serving \(v\) on the GPU
rather than from the CPU-resident graph. Then the expected benefit of
placing \(v\) on the GPU is
$
  \widehat{f}(v)\cdot \Delta t_{\mathrm{gpu}}(v).
$

Given a GPU memory budget \(B_{\mathrm{gpu}}\), selecting the
device-resident set \(R\) can be written as
\begin{equation}
\label{eq:resident_knapsack}
\max_{R \subseteq V}
\sum_{v \in R}
\widehat{f}(v)\cdot \Delta t_{\mathrm{gpu}}(v)
\quad
\text{s.t.}
\quad
\sum_{v \in R} m_{\mathrm{gpu}}(v) \le B_{\mathrm{gpu}} .
\end{equation}

Equation~\eqref{eq:resident_knapsack} can be solved as a knapsack problem. 
Exact optimization is unnecessary because it adds substantial preprocessing overhead; we therefore use a ratio-based greedy policy.
For each vertex,
{\name} computes the estimated benefit per unit of GPU memory:
\[
  \rho(v)
  =
  \frac{\widehat{f}(v)\cdot \Delta t_{\mathrm{gpu}}(v)}
       {m_{\mathrm{gpu}}(v)} .
\]
Vertices are sorted in descending order of \(\rho\), and {\name} selects them greedily until the GPU memory budget is exhausted. This ranking prioritizes vertices that combine high expected reuse with low device-memory cost. The compact sampler from Section~\ref{sec:algo} directly improves this strategy by reducing \(m_{\mathrm{gpu}}(v)\), allowing more useful sampler arrays to fit under the same memory budget.

\subsection{Update Handling}
\label{subsec:update_handling}

Graph updates are applied to the CPU-resident graph, which stores the complete graph and all sampler arrays. If an update affects a GPU-resident vertex, {\name} applies the same vertex-local update operations from Section~\ref{sec:algo} to its device copy; otherwise, no GPU-side action is required. Updates to resident and nonresident vertices can be handled concurrently by GPU kernels and CPU workers, with synchronization only before the next walk execution.

The resident set is reconsidered at epoch granularity. 
During idle periods or when the visit pattern changes substantially, {\name} reruns the selection from Section~\ref{subsec:resident_selection}. Vertices kept in the set retain their device states, newly selected vertices are copied to GPU, and unselected vertices are evicted. Only the difference between resident sets is moved.

\section{Implementation Details}
{\name} is implemented in C++ and CUDA.
%, with approximately $42$K lines of code. 
Besides standard system and CUDA libraries, our implementation uses the header-only \texttt{ankerl::unordered\_dense}  hash table~\cite{unordered_dense}. We describe its use for node2vec support in Section~\ref{subsec:algorithm-support}.

\noindent\textbf{Parameter Settings.}
We set \(c_{\mathrm{lookup}}/c_{\mathrm{find}}=0.4\) in Equation~\ref{eq:choose} for the runtime choice between rejection sampling and Fenwick-based inverse-transform sampling, as our profiling experiments show that values in the range \(0.4\)–\(1\) consistently achieve near-optimal performance.
In Equation~\ref{eq:resident_knapsack}, we use $\log d_v$ as a lightweight proxy for the GPU-side benefit $\Delta t_{\mathrm{gpu}}(v)$. 
Before the first execution round, we run a lightweight CPU-only profiling pass that accounts for $3\%$ of the total walk workload, and use it to collect initial visit statistics for resident selection.
The GPU memory budget $B_{\mathrm{gpu}}$ can be adjusted according to the available device memory in practice. 
For a fair comparison, unless otherwise specified in Section~\ref{sec:evaluation}, we set $B_{\mathrm{gpu}}$ to $20\%$ of the memory footprint required to store {\name}'s full graph and data structures.

\subsection{Dynamic Graph Support}
\label{subsec:dynamic-graph-support}

{\name} uses a lightweight dynamic graph container for edge insertions and deletions. Each vertex maintains a contiguous adjacency array with spare
capacity. When the array becomes full, {\name} doubles its capacity and copies the existing entries; under low utilization, the array is compacted and shrunk. Adjacency arrays are allocated from a memory pool to avoid
frequent small allocations.

The graph container is orthogonal to {\name}'s sampling design; more sophisticated dynamic graph stores can be used instead (Section~\ref{sec:Dynamic_Graph_Systems}), and its cost is included in all measurements and is small relative to sampler maintenance.

\subsection{Algorithm Support}
\label{subsec:algorithm-support}

{\name} separates transition sampling from workload-specific walk logic.
CPU--GPU execution proceeds in rounds. In each round, every active walker issues one sampling request. 
A workload-specific rule decides whether the sampled candidate is committed as the walker's next vertex. We increment the step count only for committed transitions:
\[
s_{walker} \leftarrow s_{walker} + \mathbf{1}[\textsc{Commit}(walker)] .
\]
Walkers that satisfy their termination condition are removed from the active set; the remaining walkers are rescheduled for the next sampling round. When the active set becomes too small to efficiently utilize the GPU, {\name} finishes the remaining walkers with a lightweight CPU-side fallback.

This interface makes workload support explicit: each workload only needs to specify how a sampled candidate is committed and when a walker terminates. We illustrate this interface using representative workloads that are also evaluated in Section~\ref{sec:overall}: DeepWalk~\cite{deepwalk}, node2vec~\cite{node2vec}, and PPR~\cite{pagerank}.

DeepWalk uses first-order random walks. Every sampled candidate is committed as the next vertex, and a walker terminates after reaching the configured walk length.

For node2vec, {\name} follows the rejection-sampling method used in KnightKing~\cite{knightking}. Each candidate transition is additionally weighted by the node2vec bias, which depends on the previous vertex $t$, the current vertex $v$, and the sampled candidate $x$:
\begin{equation}
\label{eq:node2vec}
\alpha_{pq}(t,x)=
\begin{cases}
1/p, & x=t,\\
1, & (t,x)\in E,\\
1/q, & \text{otherwise}.
\end{cases}
\end{equation}
{\name} proposes candidates from the original transition distribution of $v$ and uses the node2vec bias to decide whether a candidate is committed. We use a hash table to check whether $(t,x)\in E$. If a candidate is not committed, the walker remains at $v$, and the step count is increased only after a committed transition. Similarly, a node2vec walker terminates after reaching the walk length.

For personalized PageRank (PPR), every sampled candidate is committed as the next vertex, while termination is probabilistic. At each step, the walker terminates with probability $\gamma$; otherwise, it continues from the sampled next vertex.

\section{Evaluation}
\label{sec:evaluation}

\subsection{Evaluation Setup}
The evaluation is performed on a Linux server equipped with two 2.0 GHz Intel Xeon Gold 6330 CPUs, 56 physical cores, 512 GB of DRAM, and two NVIDIA A100 40GB GPUs. Unless otherwise stated, each experiment uses a single A100 GPU.

\noindent\textbf{Datasets.}
Table~\ref{tab:graph-stats} summarizes the statistics of our evaluation datasets. These real-world datasets are obtained from KONECT~\cite{KONECT}, the Laboratory for Web Algorithmics~\cite{WebGraph,BRSLLP,BSVLTAG} and Common Crawl~\cite{cc-domain-jan-mar-2026,cc-host-may-jul-2025}. We treat all datasets as undirected graphs, following prior systems~\cite{flowwalker,flexiwalker}. 
For unweighted graphs, we synthesize edge weights to evaluate weighted sampling. Specifically, for each edge, we generate its weight from a uniform distribution with probability $\beta$; otherwise, we assign a degree-based weight. This setting covers both random and degree-correlated weight distributions used in prior work~\cite{bingo,flowwalker,flexiwalker}. In the main experiments, we set $\beta=0.5$. We further analyze the performance of {\name} under different weight distributions in Section~\ref{sec:weight}.

To generate dynamic updates, we hold out a subset of edges from each dataset and use the remaining graph as the initial snapshot~\cite{bingo}. We then run ten update rounds; in each round, updates are randomly generated by deleting existing edges or inserting held-out edges. 
Each round contains $\mathrm{BATCHSIZE}=100\mathrm{K}$ updates.  After each round, the system generates random walks on the updated graph. We report the total end-to-end time across all ten rounds.

\begin{table}[t]
\centering
\caption{Graph statistics of the evaluated datasets.}
\label{tab:graph-stats}
\begin{tabular}{lrrrr}
\toprule
Dataset & $|V|$ & $|E|$ & Avg. Deg. & Max Deg. \\
\midrule
Arabic (AB)& 22.7M&1.3B&55.5&576.3K\\
Twitter (TW)     & 41.7M & 4.8B & 115.5 & 6.0M \\
SK-2005 (SK)     & 	50.6M& 3.9B& 76.3 & 8.6M \\
UK-Union (UK) &133.6M&11.0B&82.3& 5.6M\\
CC-Dom (CD)&120.0M&8.8B&73.7&17.8M\\
CC-Host (CH)&481.6M&6.7B&13.9&40.6M\\
\bottomrule
\vspace{-2.5em}
\end{tabular}
\end{table}

\begin{table*}[t]
\centering
\caption{Overall evaluation against four random-walk systems. Execution time is reported in seconds. Speedup is the geometric mean over completed datasets. OOM denotes out-of-memory; OOT denotes exceeding 36,000 seconds (10 hours).}

\label{tab:rw-time}
\setlength{\tabcolsep}{8pt}
\begin{tabular}{llrrrrrrr}
\toprule
Workload & System & AB & TW & SK & UK & CD & CH & Speedup \\
\midrule
\multirow{4}{*}{DeepWalk}
  & KnightKing  & 30.20 & 120.95 & 97.17 & 283.28& 317.31& 672.02 & 3.00x \\
    & FlowWalker       & 734.19& 11977.92& 5696.14& OOM & OOM & OOM & 162.57x \\
  & FlexiWalker & 47.79 & OOM & 699.48 & OOM & OOM & OOM & 8.34x\\
  & LightTraffic & 55.21 & 128.72 & 145.97 & 312.76 & 383.48 & 1048.56 & 4.05x \\
  & \name       & 13.83& 24.25 & 34.76 & 114.70 & 81.95 & 269.32&--\\
\midrule
\multirow{4}{*}{node2vec}
  & KnightKing  & 102.76 & 233.00 & 207.39 & 633.14 & 689.56 & 1613.21 &2.06x\\
    & FlowWalker      & 2079.32 & OOT & 17205.31 & OOM & OOM & OOM &74.87x\\
  & FlexiWalker & 576.44 & OOM & 5057.83 & OOM & OOM & OOM & 21.37x\\
  & \name       & 51.38 & 58.14 & 124.20 & 420.24 & 323.69 & 914.79& - \\
\midrule
\multirow{4}{*}{PPR}
  & KnightKing  & 28.04 & 74.50 & 63.76 & 214.67 & 177.74 & 229.85 &14.06x\\
    & FlowWalker     & 184.47 & 3051.01 & 2784.33 & OOM & OOM & OOM & 400.57x\\
  & FlexiWalker & 12.90 & OOM & 295.43 & OOM & OOM & OOM&26.40x \\
  & LightTraffic & 4.73 & 8.91 & 9.74 & 24.90 & 25.36 & 96.37 & 2.34x \\
  & \name       & 1.45 & 4.46 & 3.77 & 17.81 & 8.54 & 40.84 &-\\
\bottomrule
\end{tabular}
\vspace{-.4em}
\end{table*}

\noindent\textbf{Workloads.}
We evaluate three representative random-walk workloads: DeepWalk~\cite{deepwalk},
node2vec~\cite{node2vec}, and personalized PageRank (PPR)~\cite{pagerank}. Following prior
work~\cite{flowwalker,flexiwalker}, we set the walk length of DeepWalk and node2vec to $80$, use $p=2$ and $q=0.5$ in Equation~\eqref{eq:node2vec} for node2vec, and set the PPR termination probability $\gamma$ to $0.2$. 
To keep the overall evaluation tractable, we uniformly sample $10\%$ of vertices as starting vertices and use the same sources for all systems. Unless otherwise stated, the remaining sensitivity and ablation experiments use DeepWalk as the default workload.

\noindent\textbf{Baselines.}
We compare {\name} against four random-walk systems in the overall evaluation: KnightKing~\cite{knightking}, FlowWalker~\cite{flowwalker}, FlexiWalker~\cite{flexiwalker}, and LightTraffic~\cite{lighttraffic}. 
KnightKing is a representative CPU random-walk engine that uses alias sampling. FlowWalker and FlexiWalker are representative GPU-based random-walk systems. FlowWalker uses reservoir sampling, while FlexiWalker combines reservoir sampling with rejection sampling. 
LightTraffic is a representative out-of-GPU-memory design that relies on CPU--GPU transfers of graph partitions and walker states. For systems that do not natively support dynamic graphs, we make minimal implementation changes to process graph updates while preserving their original sampling algorithms.

Bingo~\cite{bingo} is a dynamic graph random-walk system that supports efficient sampling and updates, but incurs a large memory footprint. It is therefore not included in the overall evaluation because it runs out of memory on all datasets under our experimental setting. We provide a dedicated comparison with Bingo at a smaller scale in Section~\ref{sec:bingo}.

\subsection{Overall Evaluation}
\label{sec:overall}
Table~\ref{tab:rw-time} reports the execution time of {\name} and four random-walk systems across three workloads. 
Since LightTraffic does not support node2vec, we evaluate its performance only on DeepWalk and PPR.
Overall, {\name} consistently achieves the lowest execution time. Compared with KnightKing, {\name} achieves average speedups of \(3.00\times\), \(2.06\times\), and \(14.06\times\) on DeepWalk, node2vec, and PPR, respectively. KnightKing's alias table sampling is efficient for static transition distributions, but maintaining alias tables becomes expensive on dynamic graphs because any updates such as edge insertions, deletions, and weight changes can require rebuilding the affected vertex-local tables.

Compared with prior GPU-based systems, the advantage of {\name} is more pronounced. This runtime gap mainly comes from their sampling strategies rather than GPU execution itself. FlowWalker and FlexiWalker avoid heavy preprocessing structures, but their online sampling strategies become expensive on high-degree vertices or under skewed transition distributions.
In these cases, sampling may require work proportional to the vertex degree, and the largest vertices in our datasets have millions of outgoing edges.
On the datasets where they complete, {\name} outperforms FlowWalker by $162.57\times$, $74.87\times$, and $400.57\times$ on DeepWalk, node2vec, and PPR, respectively, and outperforms FlexiWalker by $8.34\times$, $21.37\times$, and $26.40\times$, respectively. 
Moreover, FlowWalker and FlexiWalker run out of memory on larger datasets, and FlowWalker exceeds the 10-hour time limit on node2vec for TW. 
These results show that GPU parallelism alone is not sufficient for large graphs; scalable random-walk execution requires an efficient sampling strategy and a memory-aware execution design.

Compared with LightTraffic, {\name} achieves average speedups of $4.05\times$ on DeepWalk and $2.34\times$ on PPR. This advantage is consistent with the CPU--GPU I/O bottleneck reported in the original LightTraffic paper~\cite{lighttraffic}, whereas {\name} keeps selected samplers resident on the GPU and transfers only walker states during walk execution.

\begin{figure}[t]
    \centering
    \includegraphics[width=.95\linewidth]{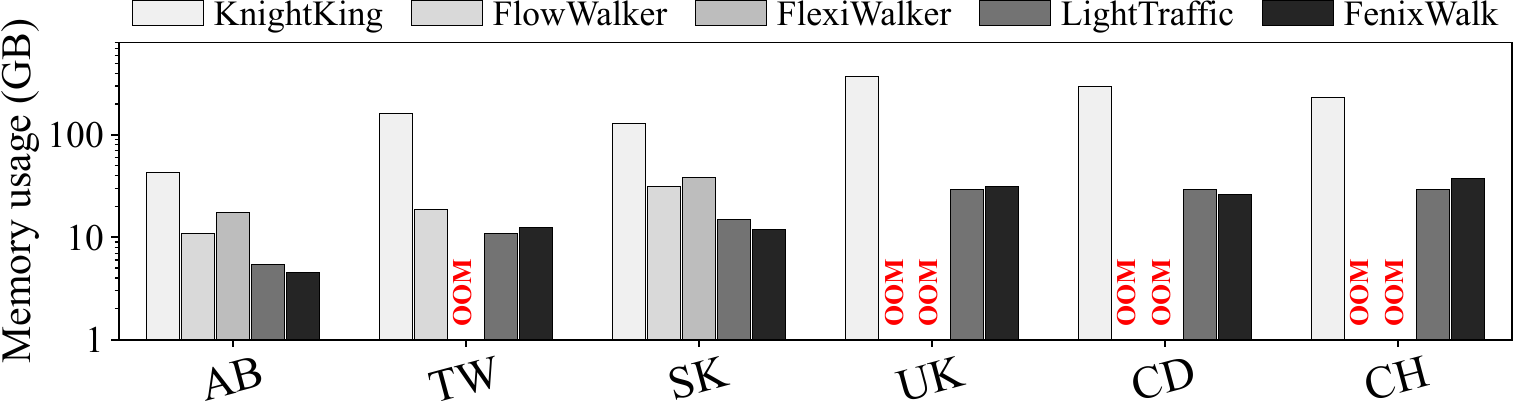}
    \caption{Memory usage in the overall evaluation.}
    \label{fig:memory}
    \vspace{-2em}
\end{figure}

The node2vec workload takes the longest time for all evaluated systems because its transition rule requires additional work beyond sampling a neighbor. In particular, each candidate step must account for the relation between the previous vertex and the candidate next vertex, which introduces extra edge-existence checks and bias computation.
%Notably, the gains of {\name} are especially large for PPR. Because PPR terminates probabilistically, many walks are short, so update and reconstruction costs are amortized over fewer sampling steps. This makes expensive sampler rebuilds particularly harmful. {\name} instead applies lightweight updates to the affected vertex-local samplers.

\noindent\textbf{Memory Usage.}
Figure~\ref{fig:memory} reports the bottleneck memory usage in the overall evaluation. We report DRAM usage for the CPU-based KnightKing and GPU memory usage for FlowWalker, FlexiWalker, LightTraffic, and {\name}. FlowWalker and FlexiWalker store only the original graph, but still require the graph to fit in limited GPU memory, leading to OOM on the larger datasets. On the datasets where they complete, {\name} also uses less GPU memory.
LightTraffic operates in the substantially simpler unbiased setting. Nevertheless, {\name} achieves comparable GPU memory usage while outperforming LightTraffic by 4.05$\times$ and 2.34$\times$ on DeepWalk and PPR, respectively. Since the reported memory resources differ, the GPU memory of {\name} is not directly comparable with KnightKing's DRAM usage. For completeness, we also measure {\name}'s CPU-side memory footprint. Across the six datasets, KnightKing consumes 42.87--231.11\,GB, compared with 28.62--204.18\,GB for {\name}, and uses $1.51\times$ as much memory on average. This saving stems from {\name}'s in-place storage of sampling information, which avoids auxiliary alias tables.

\begin{table}[t]
\centering
\caption{Time and memory comparison with Bingo on its original datasets and experimental settings.}
\label{tab:bingo_compare}
\small
\begin{tabular}{llrrr}
\toprule
Dataset & System & Time (s) & Mem. (GB) & Mem. Saving \\
\midrule
\multirow{2}{*}{AM}
  & Bingo & 0.25 & 0.87 & -- \\
  & \name & 0.33 & 0.07 & $12.43\times$ \\
\midrule
\multirow{2}{*}{GO}
  & Bingo & 0.64 & 1.38 & -- \\
  & \name & 0.49 & 0.15 & $9.20\times$ \\
\midrule
\multirow{2}{*}{CT}
  & Bingo & 0.44 & 4.62 & -- \\
  & \name & 0.40 & 0.64 & $7.22\times$ \\
\midrule
\multirow{2}{*}{LJ}
  & Bingo & 1.96 & 7.25 & -- \\
  & \name & 3.13 & 1.25 & $5.80\times$ \\
\midrule
\multicolumn{5}{c}{
  Avg. speedup over Bingo: $0.91\times$
  \quad
  Avg. memory saving: $8.32\times$
} \\
\bottomrule
\end{tabular}
\vspace{-.5em}
\end{table}

\begin{figure*}[t]
\centering

\begin{subfigure}[t]{0.34\linewidth}
\vspace{0pt}
\centering
\vspace{1.25ex}
\includegraphics[width=\linewidth]{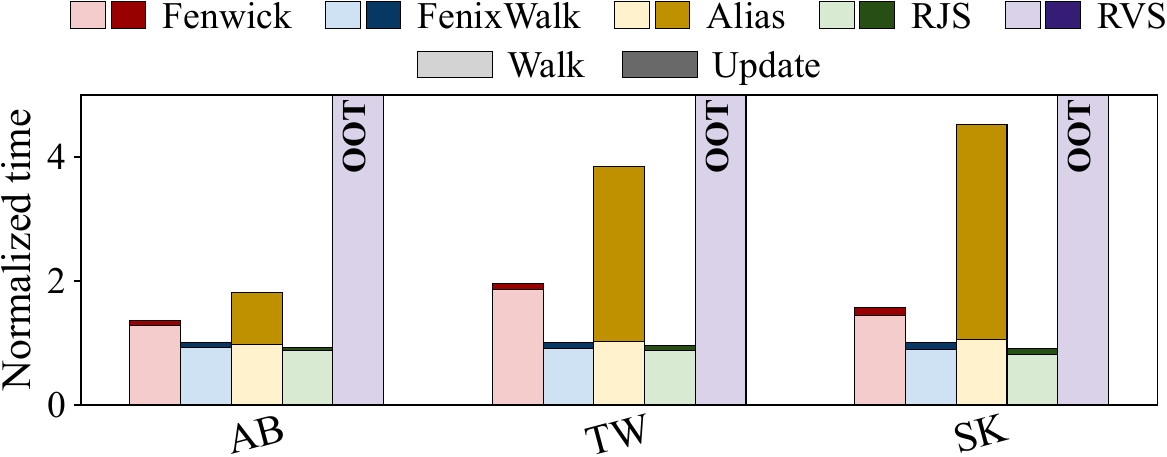}
\caption{Uniform}
\label{fig:algo-ablation-uniform}
\end{subfigure}
\hfill
\begin{subfigure}[t]{0.32\linewidth}
\vspace{0pt}
\centering
\includegraphics[width=\linewidth]{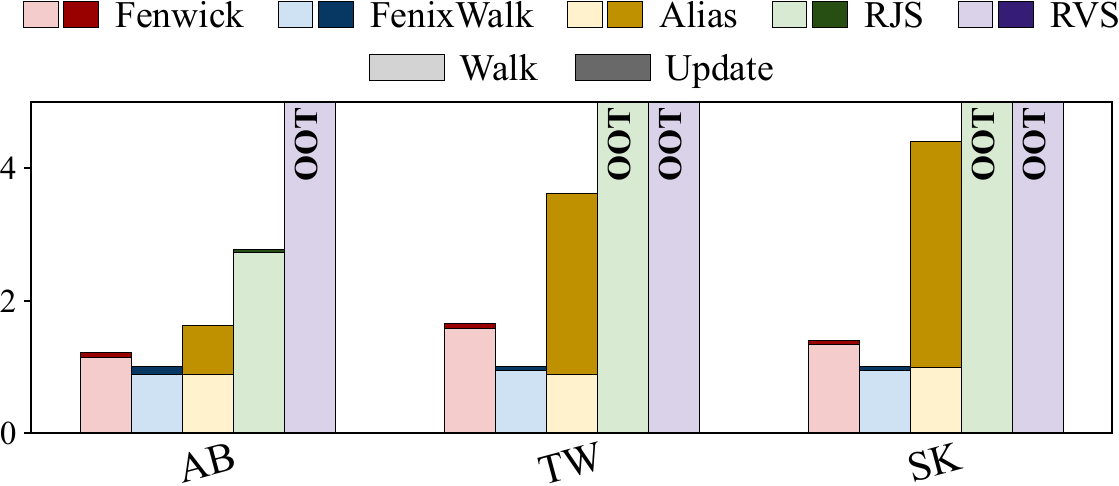}
\caption{Mixed}
\label{fig:algo-ablation-mixed}
\end{subfigure}
\hfill
\begin{subfigure}[t]{0.32\linewidth}
\vspace{0pt}
\centering
\includegraphics[width=\linewidth]{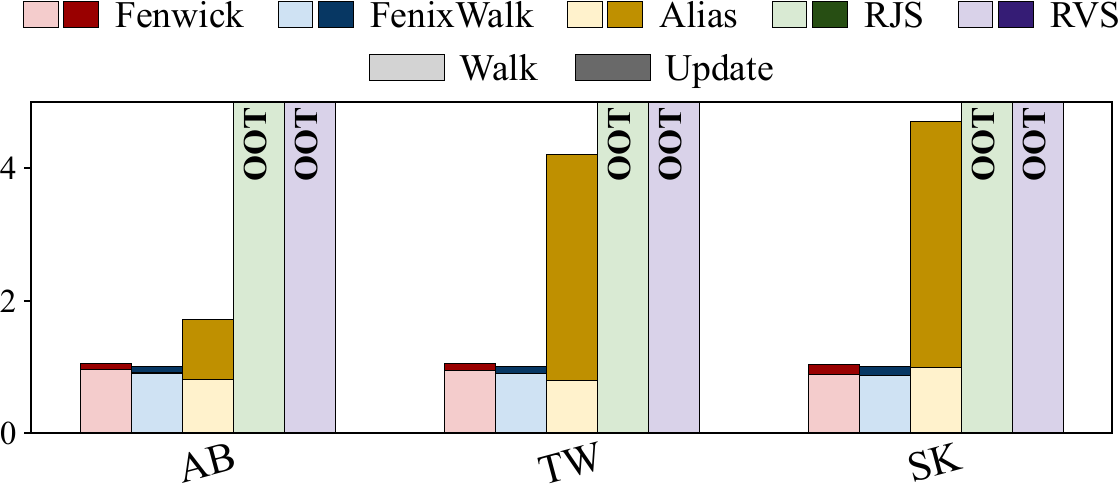}
\caption{Skewed}
\label{fig:algo-ablation-skewed}
\end{subfigure}

\caption{Algorithmic ablation under different edge-weight distributions.
We compare Fenwick-based inverse-transform sampling (\textbf{ITS}),
alias-table sampling (\textbf{Alias}), rejection sampling (\textbf{RJS}),
reservoir sampling (\textbf{RVS}), and {\name}'s sampling. Each stacked bar reports walk and update time. OOT denotes that the total time exceeds five times that of {\name}.}
\label{fig:algorithmic-ablation}
\end{figure*}

\subsubsection{Comparison with Bingo}
\label{sec:bingo}

We follow the setup of Bingo~\cite{bingo} and evaluate on the same four datasets (AM, GO, CT, and LJ) under the same DeepWalk workload. For a fair comparison, we run {\name} in the mode where all required data structures are placed on the GPU.

Table~\ref{tab:bingo_compare} shows that {\name} substantially reduces memory usage while maintaining comparable execution time under Bingo's original setting. 
Across the datasets, {\name} reduces memory consumption by $5.80\times$ to $12.43\times$, with an average saving of $8.32\times$. 
% with a geometric-mean memory saving of $8.32\times$.
In terms of runtime, {\name} is faster than Bingo on GO and CT, while Bingo is faster on AM and LJ. 
Averaged over per-dataset speedup ratios, {\name} achieves $0.91\times$ the speed of Bingo. This result is expected: Bingo's sampler provides stronger sampling complexity, while {\name} trades this advantage for a much smaller sampling footprint. At the same time, {\name} keeps update operations lightweight despite their logarithmic complexity. On an A100 GPU, {\name} reaches a peak update throughput of $1.158$ billion updates per second, compared with Bingo's $226$ million updates per second. These results suggest that {\name} removes most of Bingo's sampling-memory overhead with only minor runtime difference on small GPU-resident graphs, while retaining efficient update performance. 
Moreover, in the hybrid setting, the reduced sampler footprint allows {\name} to place more high-reuse vertex-local samplers on the GPU, leading to higher throughput under the same GPU memory budget.

\subsection{Ablation Study}

\subsubsection{Algorithmic Ablation.}
\label{sec:weight}

Figure~\ref{fig:algorithmic-ablation} compares five sampling methods under three representative edge-weight distributions: uniform-random, mixed, and skewed. The compared methods are Fenwick-based
inverse-transform sampling, alias-table sampling, rejection sampling,
reservoir sampling, and {\name}'s sampling. To isolate the sampling method from system optimizations, all variants use the same configuration and differ only in their sampler. 
Each stacked bar shows walk and update time, normalized by {\name}'s total time on the same dataset. In the uniform-random setting, edge weights are independently sampled from a uniform distribution, following prior systems~\cite{flowwalker,flexiwalker}. 
In the skewed setting, edge weights are generated based on vertex degree, following Bingo~\cite{bingo}. The mixed setting combines uniform-random and degree-based weights across vertices.

Among methods that maintain auxiliary sampling structures,
Fenwick-based inverse-transform sampling is robust to weight skew, but its total time increases to $1.35\times$--$1.96\times$
under uniform-random weights because of its logarithmic per-sample cost.
Alias-table sampling
provides efficient constant-time sampling: its walk time ranges from
$0.88\times$ to $1.18\times$ that of {\name}. However, this advantage
is offset by expensive maintenance, because updates may require
rebuilding the affected vertex-local alias table. Consequently, its
update time is up to $59.55\times$ that of {\name}, making its total
time substantially higher despite its fast sampling.

Methods without auxiliary sampling structures avoid rebuilding costs but can incur high walk-execution overhead. Rejection sampling performs well under uniform-random weights. 
Under mixed weights, however, it takes
$2.78\times$ the total time of {\name} on AB and does not complete within the time limit on the larger datasets. It also fails to complete under skewed weights because frequent rejected proposals substantially
increase its sampling cost. Reservoir sampling performs even worse because it scans the entire adjacency list for every sample. On AB, its total time is $9.48\times$ and $10.80\times$ that of {\name} under uniform-random and mixed weights, respectively, while it does not
complete within the time limit in the remaining cases. Its low update
cost therefore cannot compensate for its linear sampling overhead.

Overall, {\name} delivers the most consistent performance across weight distributions. It remains close to the fastest method under uniform-random weights and achieves the lowest time under mixed and skewed weights.

\subsubsection{System Ablation.}
\label{subsubsec:system-ablation}

\begin{figure}[t]
\centering

\begin{subfigure}{\columnwidth}
  \centering
  \includegraphics[width=\columnwidth]{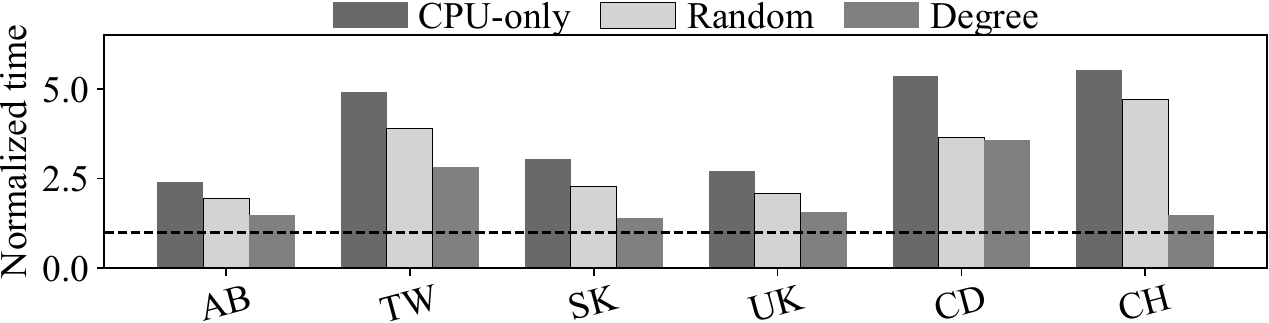}
  \caption{Normalized execution time.}
  \label{fig:resident-selection-time}
\end{subfigure}

\begin{subfigure}{\columnwidth}
  \centering
  \includegraphics[width=\columnwidth]{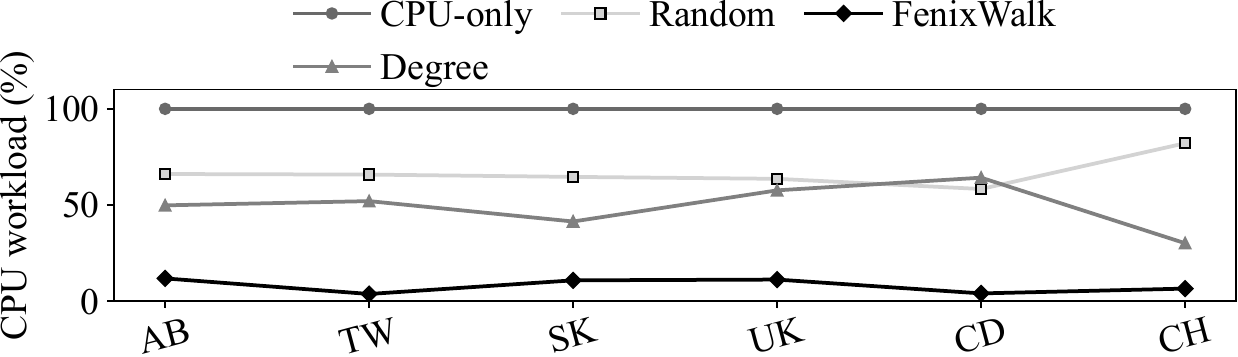}
  \caption{CPU workload ratio.}
  \label{fig:resident-selection-cpu}
\end{subfigure}
\vspace{-1.5em}

\caption{
Ablation of {\name}'s device-resident selection policy.
(a) compares execution time under different residency policies.
(b) shows the fraction of CPU workload.
}
%\vspace{-2em}
\label{fig:resident-selection-ablation}
\end{figure}

We conduct a system ablation to isolate the effects of hybrid CPU--GPU execution and device-resident selection, as shown in Figure~\ref{fig:resident-selection-ablation}. CPU-only disables GPU residency and executes all walk steps on the CPU. Random and Degree use the same GPU memory budget as {\name}, but select resident vertices uniformly at random or by highest degree, respectively.

As shown in Figure~\ref{fig:resident-selection-ablation}(a), CPU-only execution requires $2.41\times$--$5.52\times$ the execution time of {\name}, with an average slowdown of $3.99\times$. This confirms that GPU acceleration is important for random-walk execution when high-reuse vertex-local samplers can be kept resident on the device. However, simply using the same amount of GPU memory is not sufficient. Random remains $1.94\times$--$4.70\times$ slower than {\name}, because random residency does not match the skewed access pattern of random walks. 

Degree performs better than Random in execution time, reducing the average normalized time from $3.09\times$ to $2.06\times$. This shows that degree is a useful but incomplete proxy for access hotness: high-degree vertices are often more expensive to sample from and may receive more visits, but degree does not directly capture the actual walk distribution or the benefit per byte of GPU memory. 
For example, on CH, Degree leaves only $30.1\%$ of the workload on the CPU and takes only $1.50\times$ the runtime of {\name}, suggesting that degree captures a substantial part of the hot set on this graph. In contrast, on CD, Degree leaves $64.2\%$ of the workload on the CPU and takes $3.58\times$ the runtime of {\name}. This contrast shows that high-degree vertices do not always coincide with the vertices that dominate random-walk execution.

Figure~\ref{fig:resident-selection-ablation}(b) further explains the time results. Random leaves $58.3\%$--$82.1\%$ of the workload on the CPU ($66.8\%$ on average), while Degree reduces this ratio to $30.1\%$--$64.2\%$ ($49.2\%$ on average). In comparison, {\name} leaves only $3.7\%$--$11.8\%$ of the workload on the CPU ($8.0\%$ on average). Therefore, the advantage of {\name} comes not merely from using GPU memory, but from selecting samplers with high reuse relative to their device-memory cost.

\subsection{Further Analysis}

\subsubsection{Transfer Overhead}
\label{subsec:io-cost}

Table~\ref{tab:transfer-time} reports the host-to-device (H2D) and device-to-host (D2H) transfer time during execution. The measured transfer cost is small compared with the total runtime. On average, H2D and D2H account for only $4.95\%$ and $5.10\%$ of total runtime. This shows that {\name} is not dominated by CPU--GPU data movement.
Moreover, these transfer times represent the raw copy cost. In the actual execution pipeline, CPU-side sampling for nonresident walkers runs concurrently with GPU execution and data transfers. Therefore, most of the transfer latency can be overlapped with useful CPU work.

\begin{table}[t]
\centering
\small
\caption{H2D and D2H transfer time (s).}
\label{tab:transfer-time}
\begin{tabular}{lccccccc}
\toprule
Time & AB & TW & SK & UK & CD & CH & Avg. ratio \\
\midrule
H2D & 0.73 & 1.37 & 1.95 & 3.95 & 3.59 & 14.31 & 4.95\% \\
D2H & 1.15 & 1.92 & 1.36 & 2.56 & 3.45 & 10.80 & 5.10\% \\
\bottomrule
\end{tabular}
\end{table}

\subsubsection{Impact of GPU Memory Budget} 

\begin{figure}[t]
\centering
\includegraphics[width=0.95\linewidth]{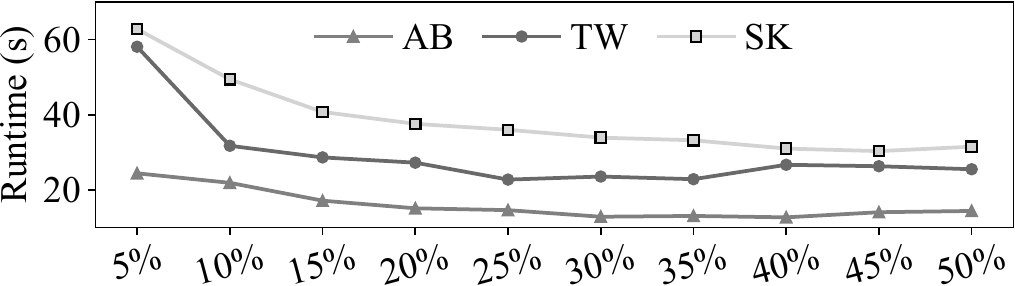}
\caption{Impact of GPU memory budget on evaluation time.}
\label{fig:memory-budget}
\end{figure}

Figure~\ref{fig:memory-budget} studies the impact of the GPU memory budget with DeepWalk workload on AB, TW, and SK. We vary $B_{\mathrm{gpu}}$ from $5\%$ to $50\%$ of the memory required to store {\name}'s full graph and data structures. Overall, increasing the GPU memory budget initially reduces the walk time, because more high-reuse vertex-local samplers can be kept on the GPU and a larger fraction of walk steps can be executed by GPU kernels.

However, the benefit is not strictly monotonic. After the budget becomes sufficiently large, increasing $B_{\mathrm{gpu}}$ provides diminishing returns and can even slightly increase the walk time. This is because a larger resident set shifts more walk steps to the GPU, and the GPU-side execution eventually becomes the bottleneck. In this regime, adding more resident vertices no longer substantially reduces CPU work, but can increase GPU kernel work. 
These results show that a moderate budget is enough to capture most of the benefit.

\subsubsection{Sensitivity to Workload and Updates}
\begin{figure}[t]
\centering
\begin{minipage}[t]{0.38\columnwidth}
\centering
\includegraphics[width=\linewidth]{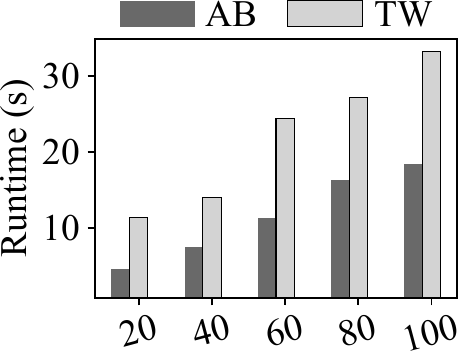}
\vspace{-0.8em}
\centerline{\small (a) Walk length}
\end{minipage}
\hfill
\begin{minipage}[t]{0.58\columnwidth}
\centering
\includegraphics[width=\linewidth]{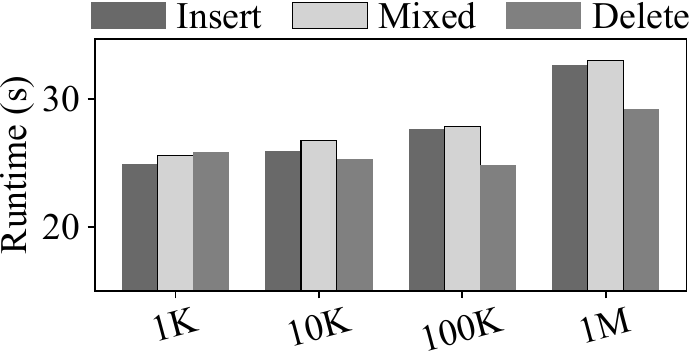}
\vspace{-0.8em}
\centerline{\small (b) Update batch size}
\end{minipage}
\caption{Sensitivity to walk length and update batch size.}
\vspace{-1em}
\label{fig:sensitivity}
\end{figure}

We further study the sensitivity of {\name} to walk length and update batch size. Figure~\ref{fig:sensitivity}(a) varies the walk length on AB and TW. As the walk length increases, the runtime increases, because longer walks generate more transition-sampling steps. This result shows that the runtime changes consistently with the amount of random-walk workload.

Figure~\ref{fig:sensitivity}(b) evaluates different numbers of updates on the Twitter graph. For insert-only and mixed updates, the runtime increases as the update batch becomes larger since more affected sampling structures need to be maintained. Delete-only updates show a different trend: the runtime first decreases and then increases. Our breakdown shows that this behavior comes from two competing effects. On one hand, more deletions increase the update-processing cost. On the other hand, deletions reduce the graph size and thus reduce the subsequent walk-execution time. The total runtime is therefore determined by the combination of increasing update cost and decreasing walk cost.

\section{Related Work}
\label{sec:related}

\subsection{Random Walk and Sampling Systems}

Random-walk systems have been studied extensively on CPUs, GPUs, and other accelerators. On CPUs, KnightKing~\cite{knightking} provides a random-walk engine with a walker-centric interface and rejection-based optimizations for node2vec. GraphWalker~\cite{graphwalker}, GraSorw~\cite{grasorw}, and SOWalker~\cite{sowalker} focus on I/O-efficient execution through subgraph loading, block scheduling, and disk-aware optimizations. ThunderRW~\cite{thunderrw} and FlashMob~\cite{flashmob} improve throughput by optimizing memory access patterns, such as step interleaving, prefetching, and cross-task reordering. TEA and TEA+~\cite{tea,tea+} use persistent data structures to support efficient temporal-graph sampling. Wharf~\cite{wharf} maintains random-walk corpora on dynamic graphs by storing walks in a compressed structure and updating only affected walk suffixes after updates.

Accelerator-based systems further exploit parallelism in random walk and graph sampling. C-SAW~\cite{c-saw} optimizes GPU-based ITS sampling, NextDoor~\cite{nextdoor} maps rejection sampling to GPUs with transit parallelism, and Skywalker~\cite{skywalker} constructs GPU alias tables, extended to multiple GPUs by Skywalker+~\cite{skywalker+}. gSampler~\cite{gsampler} provides matrix-centric APIs for sampling, while LightRW~\cite{lightrw} targets FPGA-based random-walk execution. 
LightTraffic~\cite{lighttraffic} targets out-of-GPU-memory random walks by transferring graph partitions between CPU and GPU.
Bingo~\cite{bingo} supports $O(1)$ sampling and $O(\log W)$ updates on dynamic graphs via bias decomposition.
FlowWalker~\cite{flowwalker} and FlexiWalker~\cite{flexiwalker} support random walks with runtime-dependent transition distributions; FlowWalker uses reservoir sampling, while FlexiWalker adaptively selects between reservoir and rejection sampling.

{\name} focuses on a setting that is not fully addressed by prior systems.
Most random-walk systems do not target dynamic graphs, where sampling structures may require repeated preprocessing after updates.
Wharf~\cite{wharf} instead maintains stored walk corpora under updates, whereas {\name} maintains the sampling structures used to generate walks.
Some prior work~\cite{flowwalker,flexiwalker} uses dynamic random walk to refer to state-dependent transition distributions, rather than to random walks on graphs with structural updates.
Bingo~\cite{bingo} targets dynamic graphs, but its radix-based decomposition introduces substantial memory overhead and is constrained by limited GPU memory. LightTraffic~\cite{lighttraffic} is one of the few out-of-GPU-memory designs. However, it still relies on CPU--GPU transfers of graph partitions and walk states, which are expensive for fine-grained random-walk execution. 
%Besides, its implementation only targets uniform sampling, a much simpler setting than biased sampling under graph updates.
In contrast, {\name} maintains an updateable sampler in the original weight array and uses GPU memory for high-traffic vertices without repeatedly moving large adjacency lists or auxiliary sampling structures.

\subsection{Dynamic and Streaming Graph Systems}
\label{sec:Dynamic_Graph_Systems}
Dynamic graph systems provide update-friendly storage and processing over evolving graphs.
CPU systems use specialized representations to support efficient updates, traversal, and incremental processing~\cite{stinger,GraphTinker,graphbolt,dzig,risgraph,aspen,terrace,lsgraph}.
GPU systems similarly develop dynamic storage and memory-management techniques~\cite{custinger,hornet,faimgraph,DynamicGraphsontheGPU,lpma}.
These systems are complementary to \name. They provide graph storage for dynamic processing but do not maintain sampling structures. {\name} focuses on this layer: it maintains an updateable sampler within the weight array and uses its reduced footprint to improve GPU residency in hybrid CPU--GPU execution.

\vspace{-.5em}
\section{Conclusion}
This paper presents {\name}, a hybrid CPU--GPU random-walk engine for dynamic graphs. {\name} uses a Fenwick representation to support expected $O(1)$ weight lookup for rejection sampling and $O(\log d)$ prefix sampling and updates for ITS, without retaining a separate raw-weight array. It further keeps the complete graph on the host and selectively places high-reuse vertex-local samplers on the GPU, achieving memory-efficient and scalable random-walk execution.

%\begin{acks}
%To Robert, for the bagels and explaining CMYK and color spaces.
%\end{acks}

\bibliographystyle{ACM-Reference-Format}
\bibliography{references}

%\appendix

\end{document}